\documentstyle[prl,aps,twocolumn,psfig,floats]{revtex}

\begin{document}

\title{The Different Effect of Electron-Electron Interaction on the Spectrum of Atoms and Quantum Dots\cite{footnote-1}}

\author{Kinneret Keren$^*$, Ady Stern$^*$ \& Uri Sivan$^+$ \\$^*$Department of Condensed Matter Physics\\ The Weizmann Institute of
Science\\ 
Rehovot 76100, Israel \\
$^+$Department of Physics and Solid State Institute\\
Technion-Israel Institute of Technology\\
Haifa 32000, Israel}

\date{\today}
\maketitle
\def\kf{k_{\mbox{\tiny{F}}}}
\def\rd{\rho_{\mbox{\tiny{D}}}}
\def\cf{{\mbox{\tiny{CF}}}}

\begin{abstract}

The electron-electron scattering rate of single particle excitations in atoms is estimated and compared with the corresponding rate in quantum dots. It is found that in alkali atoms single particle excitations do not acquire a width due to electron-electron interaction, 
while in complex atoms they may.
This width is typically smaller than the single particle level spacing, and hence does not affect
the number of discrete single particle excitations resolved below the
ionization threshold.
This situation is contrasted with that of quantum dots where electron-electron interaction severely
limits the number of resolved excitations.
Unlike the case of quantum dots, the scattering rate in atoms is found to decrease with increasing excitation energy. The different effect of electron-electron interaction on the spectrum of quantum dots and atoms is traced to the different confining potentials in the two systems.
\end{abstract}

\section{Introduction}

This work investigates theoretically the lifetime of single particle 
excitations due to electron-electron ({\it e-e}) interaction in atoms. In
particular we are interested in the manifestation of this lifetime 
in the measured spectrum.  
Although a large atom and a quantum dot (QD) may contain a similar number of electrons, their spectral characteristics are very different.  In QDs
the level broadening by {\it e-e} interaction 
severely limits the number of discrete single particle levels.
For example, Sivan {\it et al.}\cite{sivan} were able to resolve only $\sim 10$ discrete levels in the excitation spectrum of a diffusive 
QD containing about  $4000$ electrons. In atoms on the other hand, hundreds
of spectral
lines have been measured and tabulated, with no appreciable effect of {\it e-e} interaction on level widths. 
Hence, the effect of {\it e-e} interaction seems to be very different in these two systems, and this work attempts to understand the roots of this difference.
We do not intend to obtain 
exact results for a specific atom, but rather to develop a
general understanding of how atomic spectra are influenced by {\it e-e}
interaction in comparison with QDs.

Atoms and QDs both contain a comparable number of interacting electrons in a confined volume. However,
they differ in size, symmetry and confining potential. The latter
are typically larger, possess no particular symmetry, and are confined by a sharper potential. We find that
the different spectral characteristics of these two systems are primarily due to the different confining potentials.

We focus our attention on excited atomic states in which one electron is
excited to a weakly bound state, while the other electrons remain in their ground state configuration. The excited electron then occupies a
hydrogen-like orbital, whose spatial extent is much larger than that of the remaining electrons. The excited electron is subject to the Coulomb potential of the rest of the atom (which we refer to as ``ion''). This interaction is divided into a static 
potential (averaged over ionic degrees of freedom) and a residual interaction.  To calculate the lifetime of an excited atomic state, we consider the residual
interaction acting on the excited electron in the confining potential of the ion.

The ionic spectrum of alkali atoms is characterized by  a 
large gap (on the order of the ionization energy) at the ground state.
As a result, single particle excitations are, to 
good approximation, eigenstates of the atom, and no level broadening due 
to {\it e-e} interaction is expected. 
Complex atoms with a number of valence electrons are different.  
Typically, there are several open shells with many single particle
states of similar binding energies. 
The corresponding number of many-body states, $N_{tot}$, is hence exponentially large. We find that the interaction matrix elements are $ \propto
1/\sqrt{N_{tot}}$, while the relevant density of states is $\propto N_{tot}$.
 Consequently, for large $N_{tot}$ the matrix elements of the residual interaction are larger than the level spacing of the spectrum to which the excited state is coupled. The use of Fermi's golden rule is hence justified. The resulting level width is independent of $N_{tot}$, and typically smaller than the single-particle level spacing\cite{footnote0}.

The different effect of {\it e-e} interaction in complex atoms and QDs can be qualitatively understood in the following way. The scattering rate is proportional to the residual interaction matrix element squared times the density of final states (relaxed electron plus excited ion). In QDs, due to the sharp confining potential, the matrix elements are on the average only weakly dependent on energy. The density of final states is proportional to the excitation energy squared. The resulting scattering rates grow with energy and eventually exceed the fairly constant single particle level spacing. In complex atoms, due to the shallow confining potential, the radius of the excited electron's orbit grows rapidly with energy leading to suppressed matrix elements and diminishing {\it e-e} scattering rates. The decrease of {\it e-e} scattering rates as a function of excitation energy is in sharp contrast to the corresponding trend in QDs and Fermi liquids. The hydrogen-like single particle level spacing is also reduced with energy, but slower than the scattering rates. Consequently, in complex atoms the discrete nature of the spectrum is preserved.

The outline of this paper is as follows. In section (\ref{sec-measure}) we define
{\it e-e} lifetime of single particle excitations and discuss its meaning in a finite
system. We consider in particular the manifestation of {\it e-e} lifetime in a spectroscopic 
measurement. In section (\ref{sec-ee lifetime}) we estimate the {\it e-e}
lifetime of single particle excitations in atoms. We distinguish between alkali
atoms for which we find no broadening due to {\it e-e} interaction, and complex atoms
for which such broadening may occur.
Finally, in section (\ref{sec-con}) we compare our results for {\it e-e} lifetime in
atoms to previous results concerning {\it e-e} lifetime in QDs. 

\section{Electron-Electron Lifetime and its Manifestation in the Measured Spectrum}
\label{sec-measure}

Consider an isolated, $N$-electron system in its ground state, $ |g.s.
\rangle $,
and an excited state,
\begin{equation} \label{eq:sp excitation}
	|\Psi (t=0) \rangle = c ^{\dagger} _i c_j|g.s. \rangle ,
\end{equation}
where $c ^{\dagger} _{i} $ and $ c_j $ are single particle creation and annihilation operators respectively.
In the absence of interaction, $ |\Psi (t=0) \rangle $ is an eigenstate, and 
therefore characterized by an infinite lifetime. In the presence of interaction, it is no longer an eigenstate and decays with time. 
The {\it e-e} lifetime, denoted by
$ \tau _{ee}$, is defined as the decay time of the initial state,
\begin{equation} \label{eq:sp lifetime}
	| \langle \Psi (t) |\Psi (t=0) \rangle | ^{2} \sim e^
	{- \frac {t}{\tau _{ee}}} .
\end{equation}
This analysis assumes zero temperature. Generalization to finite temperature is straightforward. 

The meaning of {\it e-e} lifetime in a system characterized by a discrete energy spectrum, 
such as an atom below the ionization threshold or a QD, should be clarified. Let $ 
\{|\psi_{\alpha} \rangle\} $ be a basis of exact eigenstates
of the system with energies $  \{E_{\alpha} \} $.
The initial state can be expressed as a superposition of these eigenstates,
\begin{eqnarray} \label{eq:decomposition}
	& &|\Psi (t=0) \rangle = c ^{\dagger} _i c_j |g.s. \rangle =
	\sum_{\alpha} \lambda_{\alpha} |\psi_{\alpha}\rangle \\
	& &\sum_{\alpha} |\lambda_{\alpha}|^2=1. \nonumber
\end{eqnarray}
The time evolution of the initial state is 
given by,
\begin{equation}
	|\Psi (t) \rangle = 
	\sum_{\alpha} \lambda_{\alpha} e^{i E_{\alpha} t/\hbar} |\psi_{\alpha}
	\rangle,
\end{equation}
and the probability to remain in the initial state at time $t$ is,
\begin{equation}
	|\langle \Psi(t=0)|\Psi (t) \rangle |^2= 
	\left|\sum_{\alpha} |\lambda_{\alpha}|^2 e^{i E_{\alpha} t/\hbar}
	\right|^2.
\end{equation}
Strictly speaking, as opposed to the case of a continuous spectrum, here there
is no decay of the probability to remain in the initial state.
This probability oscillates, and for long enough times it can get arbitrarily 
close to $1$ (it can return exactly to $1$ if the energy spectrum is commensurate).
Nevertheless, as the number of dominant terms in (\ref{eq:decomposition})
increases, it takes longer for the initial state to reconstruct. The typical time scale for this reconstruction is the Rabi-time, $\tau_{Rabi}=\hbar/\Delta$, where $\Delta$ is the average many-body level spacing. On shorter time scales the probability effectively decays exponentially with time. Consequently, we define the {\it e-e} lifetime in a finite system
as in (\ref{eq:sp lifetime}), keeping in mind the restriction on the time 
scales $t,\tau_{ee} \ll \tau_{Rabi}$.

The {\it e-e} lifetime of a single particle excitation can be
estimated using Fermi's golden rule.
The full interacting Hamiltonian $H$ is divided into an unperturbed Hamiltonian
$H_0$, which may include part of the e-e interaction (for example, the Hartree
or Hartree-Fock part), and a perturbation   
Hamiltonian $H_{int}$ which includes the rest of 
the {\it e-e} interaction. Let $ \{|f^0 \rangle \}$ be the $N$-particle eigenstates 
of $H_0$ with eigenenergies $\{ E_f^0 \}$. 
The golden rule result for the {\it e-e} lifetime of a single particle excitation
$|i^0 \rangle$  is,
\begin{equation} \label{eq:GR}
	\tau_{ee}^{-1}=\frac{2 \pi}{\hbar} \sum_{f\neq i} 
	\left|\langle f^0|H_{int}|i^0 \rangle \right|^2
	\delta(E_i^0-E_f^0).
\end{equation} 
The sum in (\ref{eq:GR}) extends over all N-particle eigenstates of $H_0$
other than the initial one. 

Following \cite{AAGK}, we arrange the eigenstates of the unperturbed Hamiltonian $H_0$ in the form of a
hierarchal tree emanating from the ground state of $H_0$. The first generation
includes all eigenstates of $H_0$ which are coupled by the interaction to the ground state, in particular single particle excitations (as in (\ref{eq:sp excitation})) are included. Each generation
includes all eigenstates of $H_0$ not previously incorporated in the tree which
are connected by non-vanishing interaction matrix elements to the previous
generation. The problem of {\it e-e} lifetime of single particle excitations can be mapped on the problem of Anderson localization on the hierarchal tree, described by a Hamiltonian $H_{A}$. 
Each eigenstate $|f^0\rangle$ of $H_0$ is a site in the hierarchal tree with an on-site energy $E^0_f$.
The hopping amplitude between two sites is given by
the corresponding residual interaction matrix element. The single particle spectrum of  $H_{A}$ on the hierarchal tree is equivalent to the many-body spectrum of $H_0$. 
An excitation of the type (\ref{eq:sp excitation}) corresponds to a single site on the lattice. Its time evolution is determined by the overlap of this site with exact eigenstates of $H_A$. An overlap of this site with extended states leads to exponential decay, while overlap limited to localized states leads to beating of few frequencies. 

 We now turn to discuss how {\it e-e} lifetime of single particle
excitations is manifested in the measured many-electron spectrum of a quantum dot or an atom.
A finite, isolated, Fermi system has discrete energy levels of zero width, which
correspond to {\em exact} many-body eigenstates of the system. 
A spectroscopic measurement 
involves coupling the system to some external measuring device. 
Typically, the measurement operators are single electron ones. As a
result, only many-body eigenstates with finite overlap with single particle
excitations can be detected.

Consider a specific example, an optical
absorption experiment on a many-electron system in its ground state 
$|g.s. \rangle$. The measurement operator in this case is given by,
\begin{equation}
	O=\sum_{i,j} \lambda_{ij} c^{\dagger} _i c _j + h.c. \:,
\end{equation}
where $\lambda_{ij}$ are coefficients (including matrix elements of photon operators) determining the
strength of the coupling to various single particle excitations.
The absorption of the system
is characterized by the spectral function,
\begin{eqnarray}
	\alpha(\omega)  = 
	\sum_f & | \langle f|\sum _{i,j} \lambda_{ij} c
	^{\dagger} _i c _j + h.c.|g.s. \rangle |^2
	\\ \nonumber &   \times 
	\delta(E_{gs}+\hbar \omega-E_f),
\end{eqnarray}
where $|f\rangle$ and $E_f$ are exact final many-body eigenstates and 
eigenenergies of the system, and $E_{gs}$ is the ground state energy of the system. 
In the absence of {\it e-e} interaction, $\alpha(\omega)$ exhibits a series of
 $\delta$-peaks
corresponding to single particle excitations of the system.
 When {\it e-e} interaction is included,
each single particle excitation becomes a superposition of several exact many-body eigenstates. The spectral function then displays many more 
absorption peaks corresponding
to many-body eigenstates that overlap with single particle excitations. 
A many-body eigenstate can overlap several single particle excitations, 
which then generate interfering contributions to its absorption peak.

A significant simplification of the spectral function and its  relation to the
concept of {\it e-e} scattering rate is obtained by assuming that each many-body state $|f \rangle$ overlaps with at most one single particle excitation, an
approximation equivalent to the reduction of the hierarchal tree into a Cayley tree\cite{AAGK}. This
approximation is valid when the
{\it intra}-generation matrix elements are negligible, or when the single particle
level spacing is much larger than the resulting {\it e-e} broadening. Within 
this approximation,
the intensity of peaks having significant overlap with a particular
single-particle excitation $c^{\dagger}_{i_0} c_{j_0}|g.s. \rangle$ is approximated by $|\lambda_{i_0 j_0}|^2 |\langle f| c^{\dagger} _{i_0} c _{j_0} |g.s. \rangle|^2$.
The relative intensity of all
$\delta$-peaks $|f\rangle$ associated with this single particle excitation  is
proportional
to $|\langle f| c^{\dagger} _{i_0} c _{j_0} |g.s. \rangle|^2$. The factor
$|\lambda_{i_0 j_0}|^2$ is common to all of them.

Consider the time evolution of a system initially in a state,
$|\psi(t=0) \rangle = \; c^{\dagger}_{i_0} c_{j_0} |g.s.\rangle$,
\begin{equation}
        |\psi(t) \rangle=\sum_f e^{i E_f t/\hbar}
        \langle f|c^{\dagger}_{i_0} c_{j_0} |g.s.\rangle |f\rangle.
\label{eq:time}
\end{equation}
The {\it e-e} scattering rate is determined by the
energy spread of exact many-body eigenstates, $|f\rangle$, that participate in the sum (\ref{eq:time}), provided there are many such states. 
These many-body eigenstates are exactly those that generate the absorption peaks associated with  $c^{\dagger}_{i_0} c_{j_0}|g.s.\rangle$ in the spectral function. Thus, {\it e-e} scattering rate of a single particle excitation is manifested in the 
energy width of the cluster of absorption peaks associated with it in the spectral function.

In a real experiment the resolution of the measuring device is finite. 
The measured  many-body spectrum
is smeared, so that each $\delta$-peak in the spectral function appears as a
broadened peak.
This broadening may have a typical
scale larger than the many-body level spacing. 
In this case, the many-body eigenstates can no longer be resolved in the
spectrum. Then, a cluster of $\delta$-peaks associated with a given single particle excitation appears as a broad single particle resonance 
whose width is equal to the inverse {\it e-e} lifetime of the excitation.
Single particle resonances can be resolved as long as their broadening is smaller than the 
single particle level spacing. When the broadening of single-particle resonances  
exceeds the single-particle level spacing, the
measured spectrum becomes essentially continuous.

\section{{\it e-e} Lifetime in Atoms} \label{sec-ee lifetime}

In this section we investigate the {\it e-e} lifetime of single particle excitations
in atoms or ions below the ionization threshold\cite{footnote1}.
We consider an atom (or ion) composed of a fixed nucleus  with charge $Z$ and  $N$ electrons.
The full non-relativistic Hamiltonian of such a system is,
\begin{equation} \label{eq:full atomic Hamiltonian}
	H=\sum_i (\frac{p_i^2}{2 m}-\frac{Z e^2}{r_i})+\sum_{i \neq j}
	\frac{e^2}{|\vec{r_i}-\vec{r_j}|}\; ,
\end{equation}
where $ \vec{p}_i $ and $ \vec{r}_i $ are the momentum and position operators of
the electrons, respectively, $m$ is the electron's mass and $e$ is the electron's
charge. Since an exact solution to the full Hamiltonian is generally not known, one resorts to approximation methods. A standard approximation is the replacement of the full
Hamiltonian by a single particle Hamiltonian that includes an effective potential
induced by the average electron density (e.g. the Hartree approximation). 
The resulting single particle spectrum is degenerate due to spherical symmetry, leading to the well known shell structure of atoms.

The degeneracy in the single particle spectrum leads to a much
larger degeneracy in the many-body spectrum. For example, the ground state 
configuration of an Europium atom includes $7$ electrons in an open f-shell. 
The $14$-fold single particle degeneracy of an f-shell, yields a 
$14!/(7!)^2=3432 $-fold degeneracy in the 
non-interacting many-body ground state.  Within such degenerate subspaces
the interaction matrix elements are in many cases comparable or larger than
the many-body level spacing. As a result, {\it e-e} interaction has significant
effect on the many-body eigenstates and eigenenergies. This is demonstrated in the numerical calculations of the eigenstates and eigenenergies of a Cerium atom by Flambaum {\it et al.}\cite{Flambaum}.
Thus, in a perturbative calculation of {\it e-e} lifetime  most of the interaction has to be included in the unperturbed Hamiltonian, 
$H_0$, which is no longer a simple sum of single particle Hamiltonians.

We consider the {\it e-e} lifetime of high energy  
single particle excitations (but still below the ionization threshold). As a result the excited electron spends most of the time
in regions where the density of the other electrons is exponentially
small. Two major simplifications can hence be made,
\begin{enumerate}
	\item The exchange integral between the excited electron and the rest of the electrons is exponentially small and can be neglected. The excited electron can therefore be considered distinguishable from other electrons.
	 \item The excited electron is subject to a potential which is roughly that of a hydrogen atom. Its wavefunction may then be approximated by the corresponding hydrogenic one, except for s-shell electrons that penetrate into the ion.
\end{enumerate}

The atom is divided in our treatment into an excited (distinguishable) electron and an ion containing all other electrons and the nucleus.  
The Hilbert space is spanned by direct products of
the ionic states and the excited electron states.

The unperturbed Hamiltonian is,
\begin{eqnarray} 
	H_0=H_{ion}+h_{e}\; ; & \;\;\; & h_e=\frac{p^2}{2m}+V_e(r).
\end{eqnarray}
$H_{ion}$ is the full Hamiltonian of the ion, and $h_e$ is the effective single particle Hamiltonian of the
excited electron which includes an effective
potential $V_e$ induced by the nucleus and the spherically averaged ionic ground state electron 
density. In regions exterior to the ionic electron density,  
the effective potential experienced by the excited electron is simply
$V_e(r)\simeq-e^2/r$.
The perturbation Hamiltonian is,
\begin{eqnarray} \label{eq:new Hint}
	H_{int}&=&e^2 \int d\vec{r'}\frac{\delta\rho(\vec{r'})}
	{|\vec{r}-\vec{r'}|} \\
	\delta\rho(\vec{r'})&=& \rho(\vec{r'})-\bar{\rho}(\vec{r'}) \\
	\nonumber & = &
	\rho(\vec{r'})-\frac{1}{4\pi}\int d\Omega\langle g.s|
	\rho(\vec{r'})|g.s.\rangle, 
\end{eqnarray} 
where $\rho(\vec{r'})$ is the density operator of the $N-1$ 
electrons of the ion, and
$\delta\rho(\vec{r'})$ is its fluctuating part. $|g.s \rangle$ refers to the
ionic ground state, and $\vec{r}$ is the position operator of the excited electron.

The golden rule is now employed to the calculation of the lifetime of an excited electron due to the residual coupling $H_{int} $ with the rest of the electrons. We first consider the relevant density of states and the interaction matrix
elements.

\subsection{The Density of States of $H_0$}

The density of states of $H_0$, $G(E)$, is a convolution of the density of
states of the excited electron, $g_e(\epsilon)$, and the density of states of the
ion, $g_{ion}(\epsilon)$,
\begin{equation}
	G(E)=\int d\epsilon \: g_e(\epsilon) g_{ion}(E-\epsilon),
\end{equation}
where all energies are measured relative to the ground state.
In order to study the spectrum of $H_0$, we first examine the single 
particle spectrum of the atom within the Thomas-Fermi 
model, which gives a fairly good picture of a large enough atom. We later
consider the characteristics of the many-body spectrum of the ion and 
the effect of {\it e-e} interaction. 

\subsubsection{The Single Particle Spectrum within the Thomas-Fermi Model}
\label{sec-sp TF}

The Thomas-Fermi model (TF) (see e.g. \cite{LL}) is a self-consistent single particle model 
 used to obtain the approximate ground state electron density of an atom from
which an effective single particle potential is calculated. Here, we use 
it to estimate the density of states for excitations above the atomic ground state. The TF effective potential in a neutral atom is the solution of the TF equation,
\begin{equation} \label{eq:TF}
	 \frac{d^2 V(r)}{dr^2}=-\frac{8 \sqrt{2}}{3 \pi} \frac{1}{e a_0^{3/2}} 
	(\mu-V(r))^{3/2},
\end{equation}
with,
\begin{eqnarray}
	V(r \rightarrow 0) \sim - \frac{Z e^2}{r} 
	 \;\; ; & \;\;\; &  V(r \rightarrow \infty) \rightarrow 0. 
\end{eqnarray}
Here, $\mu$ is the chemical potential and $a_0=\hbar^2/me^2$ is the Bohr radius.
We define, as customary,
\begin{eqnarray} 
	& & x=\frac{r}{a_0} Z^{1/3} b \;\; ; \;\;\;
	b=2 \left(\frac{4}{3 \pi} \right)^{2/3} 
	\label{eq:x} \\ 
	& &\chi(x) =  - \frac{V(r)-\mu}{Z e^2 /r}.
\end{eqnarray}
With these definitions, the TF equation reduces to, 
\begin{eqnarray} \label{eq: chi for neutral atom}
	 & & x^{1/2} \frac{d^2 \chi}{dx^2}=\chi^{3/2}(x)  \\ \nonumber
	 & &\chi(0)=1 \;\; ;  \;\;\;  \chi(x \rightarrow \infty)=0.
\end{eqnarray}
This non-linear differential equation is solved numerically\cite{LL}, 
and  we denote its solution by $\chi_0(x)$. Since 
the equation and its boundary conditions are independent of $Z$, so must be its solution $\chi_0(x)$. The TF potential is given by,
\begin{equation}
	V_{TF}(r)=-\frac{Z e^2}{r} \chi_0 (\frac{r}{a_0} Z^{1/3} b) .
\end{equation}
Asymptotically,
\begin{eqnarray} 
	V_{TF}(r) \sim \left\{ \begin{array}{ll}
		-  \frac{Ze^2}{r} & r \ll Z^{-1/3}a_0 \\
		-  \frac{144}{b^3} \frac{e^2 a_0^3}{r^4} & r \gg Z^{-1/3}a_0.
		\end{array} \right.
\end{eqnarray}
The length scale 
$ Z^{-1/3} a_0 $ characterizes screening in a TF atom.

An obvious flaw in the TF potential is that it includes the electrostatic 
self-interaction of the electron. This has a noticeable effect at large 
distances where the electron is essentially outside the ionic
charge distribution, and therefore experiences a potential $V(r) \simeq -e^2/r$, rather than $ \sim 1/r^4 $ as obtained from the TF model.
The TF potential can be modified to correct this erroneous result in various ways\cite{Latter} leading to,
\begin{eqnarray} 
	\tilde{V}_{TF}(r) \sim \left\{ \begin{array}{ll}
		-  \frac{Ze^2}{r} & r \ll Z^{-1/3}a_0 \\
		-  \frac{144}{b^3} \frac{e^2 a_0^3}{r^4} & Z^{-1/3}a_0 \ll r
		< a_0 \\
		- \frac{e^2}{r} & r \gg a_0 .
		\end{array} \right.
\end{eqnarray}
The modification to the potential is important for the excited
electron we consider, since it spends most of its time at distances larger than $a_0$ from the nucleus.

The ionization energy of the atom is the
difference between the chemical potential of a neutral atom ($\mu=0$), and
that of an ion ($N=Z-1$).
To lowest order in $1/Z$, the ionization energy is
constant and given by\cite{March}, 
\begin{equation} \label{eq:ionization energy}
	E_{ion} \approx 0.109 \frac{e^2}{a_0}.
\end{equation} 
The ionization energy is approximately the
same for all atoms. This result does not account for the irregular variation
of $E_{ion}$ as a
function of $Z$ arising from the shell structure.

The single particle density of states of a TF atom is obtained 
 by numerically solving  the Schr\"{o}edinger equation with $V_{TF}$.
Alternatively,
the single particle density of states can be calculated in the
semiclassical approximation employed in the
derivation of  the TF equation. Since we are interested only in the approximated density of states\cite{footnote2}, the semiclassical approach is
sufficient. 

The density of states of a free 3D Fermi gas enclosed in a volume $V$ is given by  $g(\epsilon)=(2m)^{3/2}/(2 \pi^2 \hbar^3) \sqrt{\epsilon} \: V$. 
To obtain the  density of states in a TF atom with a space dependent
potential 
we use the expression for the free Fermi gas locally, replacing $\epsilon$ by 
$\epsilon-V_{TF}(r)$ at each point. Integrating over space we obtain
the single particle density of states in the semiclassical 
approximation,
\begin{equation} \label{eq:TF dos}
	g_{TF}(\epsilon)=2\frac{(2m)^{3/2}}{\pi \hbar^3} \int_0^{R(\epsilon)}
	r^2 \sqrt{(\epsilon-V_{TF}(r))} \: dr. 
\end{equation}
where $R(\epsilon)$ is the classically accessible radius given by
$V_{TF}(R(\epsilon))=\epsilon$. 
	
Introducing explicit Z-dependence and using the definition of $x$ in (\ref{eq:x}) we obtain,
\begin{eqnarray} \label{eq:gtf(e,z)}
 	\lefteqn{g_{TF}(\epsilon,Z)} & & \\ \nonumber
	 & = & 2\frac{(2m)^{3/2}}{\pi \hbar^3} 
 	\int_0^{R(\epsilon,Z)} r^2 \sqrt{(\epsilon-V_{TF}(r,Z))} dr \\
	\nonumber
 	& = & \frac{1}{\pi} \left( \frac{3 \pi}{4} \right)^{5/3} Z^{-1/3} 
	\frac{a_0}{e^2}
 	\\  & & \times \int_0^{X(\epsilon a_0/Z^{4/3}e^2)}
 	x^2 \sqrt{\frac{\epsilon a_0}{Z^{4/3} b e^2} +\frac{\chi_0(x)}{x}}  dx
	\nonumber.
\end{eqnarray}
$X=R(\epsilon,Z) Z^{1/3}b/a_0$ is
the solution of $\chi_0(x)/x=-\epsilon a_0/b Z^{4/3} e^2$, hence $X=X(\epsilon
a_0/Z^{4/3}e^2)$. 
We find that $g_{TF}(\epsilon,Z)$ 
obeys the following scaling rule,
\begin{equation}
	g_{TF}(\epsilon,Z)=Z^{-1/3} \frac{a_0}{e^2}\;
	f \left( \frac{\epsilon a_0}{Z^{4/3}e^2} \right),
\end{equation}
where  $f$ stands for the integral appearing  in (\ref{eq:gtf(e,z)}).

Consider the single particle density of states at energies comparable to the ionization energy,
$\epsilon \sim -E_{ion}$, since only electrons in 
this energy range  participate in real scattering processes
contributing to the {\it e-e} lifetime of excitations below the ionization threshold.
We refer to such electrons as {\it active} electrons. In order to evaluate the TF single particle density of states, $g_{TF}(\epsilon,Z)$, for  $\epsilon \sim -e^2/a_0$,
we divide the spatial integration in (\ref{eq:TF dos}) into two parts:
\begin{description}
	\item[a.] {\bf Core ($r<10\; Z^{-1/3}a_0$)} - In this region 
	$V_{TF}(r) \sim Ze^2/r$,
	and $|\epsilon|\ll|V_{TF}(r)|$. The contribution of this region to
	$g_{TF}$ can thus be approximated by,
	\begin{eqnarray}
		g_1(\epsilon) & \sim & 2\frac{(2m)^{3/2}}{\pi \hbar^3} 
		\int_0^{10 Z^{-1/3} a_0}
		r^2 \sqrt{\frac{Z e^2}{r}} dr \\ \nonumber
		& \sim & Z^{-1/3} \frac{a_0}{e^2}.
	\end{eqnarray}
	The physical meaning of this is that for large $Z$
	the core electrons lie deep in the atomic potential well, and their
	contribution to the single particle density of states at high energies
	is negligible.
	\item[b.] {\bf Outer shell ($10\; Z^{-1/3}a_0<r$)} -  Substituting the
	asymptotic expression for the potential, $V_{TF}(r)\simeq -144 e^2
	a_0^3/b^3 r^4 $, valid for $r\gg Z^{-1/3} a_0$, we obtain, 
	\begin{eqnarray}
		g_2(\epsilon) & < & 2\frac{(2m)^{3/2}}{\pi \hbar^3} 
		\int_{10 Z^{-\frac{1}{3}} a_0}^{R(\epsilon)}
		r^2 \sqrt{\epsilon+\frac{144 e^2 a_0^3}{b^3 r^4}} dr 
		\\ \nonumber & \simeq & 50
		\frac{a_0}{e^2} \left( \frac{e^2}{\epsilon a_0} \right)^{1/4}.
	\end{eqnarray}		
\end{description}

Thus, the TF model indicates that most electrons belong to the core, and only a small number
of them ($\ll Z$) are active.
For a neutral atom the number of active 
electrons is estimated by,
\begin{equation} \label{eq:nefftf}
	N_{eff}^{TF}(Z) = \int_{-e^2/2 a_0}^0 g_{TF}(\epsilon,Z) d\epsilon.
\end{equation}
The contribution of the core electrons to $N_{eff}^{TF}(Z)$  is $\sim Z^{-1/3}$, so for large atoms $N_{eff}^{TF}$ is composed solely of electrons from the outer shell.

\begin{figure}[!hbt]
	\centerline{\psfig{figure=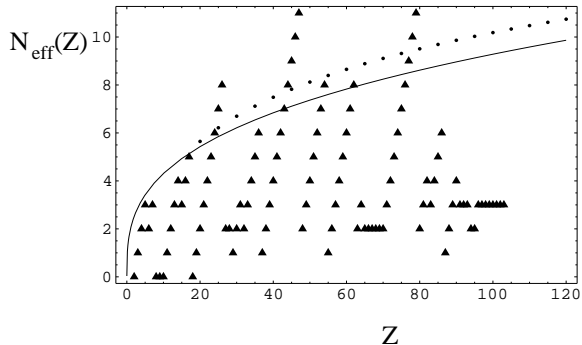,width=3in}}
	\caption{Dotted line: $N_{eff}^{TF}(Z)$ as a function of $Z$ (eq.\protect\ref{eq:nefftf}). Solid line: $2\cdot Z^{1/3}$. Triangles: $ N_{eff}(Z)$ as calculated from the discrete single particle spectrum.}
	\label{fig-neff(Z)}
\end{figure}

The results of numerical calculation of $N_{eff}^{TF}(Z)$ are
shown in figure  (\ref{fig-neff(Z)}), for $Z$ values  characteristic of  real atoms.
Indeed $N_{eff}^{TF}(Z) \ll Z$.
In this range ($Z<100$), $N_{eff}^{TF}$ is found to be roughly proportional
to $Z^{1/3}$. However,  
within the TF model, $N_{eff}^{TF}(Z)$ does not diverge as $Z \rightarrow
\infty$. Rather it is upper bounded\cite{footnote3} by,
\begin{eqnarray} 
	\forall \:Z & \;\;\; & N_{eff}^{TF}(Z)<40.
\end{eqnarray}
For large enough $Z$ the number of active electrons in a TF atom is
constant, independent of $Z$. There are, however, no real atoms with
large enough $Z$ to check this prediction experimentally. 

For a particular atom, the actual number of active electrons,
$N_{eff}(Z)$, depends irregularly on $Z$ due to the shell structure. The number of electrons in
the ground state configuration with single particle energy $\epsilon>-e^2/2a_0$, according to a Hartree-Fock-Slater calculation\cite{atomic calc}, is compared with $N_{eff}^{TF}(Z)$ in figure
(\ref{fig-neff(Z)}). It is $N_{eff}(Z)$ and not
$N_{eff}^{TF}(Z)$, which
determines the spectra of different atoms. Nevertheless,
$N_{eff}^{TF}(Z)$ gives a rough estimate for $N_{eff}(Z)$, 
and leads us to the important observation that this number is $\ll Z$.

\subsubsection{The Spectrum of $H_0$}

The TF model suggests that the single particle density of states of an atom
(or ion) is finite at the atomic Fermi
energy, even for
 $Z \rightarrow \infty$. However, because of the degeneracy of
the single particle spectrum and the importance of {\it e-e} interaction,
the characteristics of the many-body spectrum vary strongly
between atoms. We recall that the density of states of $H_0$ is a convolution 
of the density of states of the excited electron, $g_e$, and the ionic many-body density of states, $g_{ion}$. Apart from the few lowest levels, the spectrum of the excited electron is hydrogen-like in all atoms. Hence, $g_e$ depends only weakly on the specific atom under consideration. 
The converse is true for the 
ionic density of states since the many-body density of states in the
vicinity of the ground state strongly depends on the number of valence 
electrons and on
the number of available single particle states of similar energy.
We discuss two limiting cases - alkali atoms characterized by a sparse 
spectrum in the vicinity of the ground state, and complex atoms 
characterized by an exponentially large density of states in that energy range.

Alkali atoms are characterized by a single valence electron in an s-shell. 
All other electrons reside in closed shells with relatively
deep single particle energies. Exciting an additional 
electron has a relatively large energy cost, hence the spectra of
singly ionized ions have large gaps (nearly $\sim E_i$) at the 
ground state. 

Atoms with a few valence electrons are termed ``complex atoms''. Typically
such atoms have several open shells with similar single particle  energies in
the vicinity of the Fermi level. As $Z$ increases these shells can acquire
large orbital momenta, and consequently large degeneracy. The distribution of 
$N_{eff}-1$ valence electrons of the ion among those $N_s$ single particle states gives rise to  $\sim N_s^{N_{eff}-1}$ many-body states of comparable energy, leading to an average many-body level spacing,  
$\Delta \sim N_s ^{-(N_{eff}-1)} e^2/a_0$. 

The effect of {\it e-e} interaction on the many-body spectrum of a complex atom is
demonstrated in the numerical calculation of the spectrum and eigenstates of a
Cerium atom by Flambaum {\it et al.}\cite{Flambaum}. A Cerium atom contains only $4$ valence
electrons, which are sufficient to generate a complex many-body spectrum.
The eigenstates of a Cerium atom are shown to become chaotic superpositions of
Slater determinants belonging to different configurations, already at low
excitation energies.
It is evident that {\it e-e} interaction can not be treated perturbatively in complex atoms. Generally (a) it lifts most degeneracies peculiar to the non-interacting spectrum, leading to a more homogeneous spectrum and (b) it mixes Slater determinants, so that exact eigenstates are superpositions of many determinants. The single particle selection rules are hence relaxed, and only rules concerning the total spin and angular momentum hold. Consequently, the density of states for transitions is significantly increased.

\subsection{The Interaction Matrix Elements}
\label{sec-me interaction}

We consider matrix elements of $H_{int}$ between the initial 
excited state, $ |i \rangle=|n_i,l_i,m_i \rangle |\alpha_i \rangle $,
and any final state, $ |f \rangle=|n_f,l_f,m_f \rangle |\alpha_f \rangle $. 
$|n_i,l_i,m_i \rangle$ and $|n_f,l_f,m_f \rangle$  refer, respectively, to the initial (hydrogen-like) and final state of the excited
electron\cite{footnote4}. Similarly,  $|\alpha_i \rangle$ and $|\alpha_f \rangle$ refer to the initial and final state of the ion.
$|\alpha_i \rangle$ and $|\alpha_f \rangle$ are exact eigenstates of the ion,
so that $|i \rangle$ and $|f \rangle $ are eigenstates of $H_0$. 

The multipole expansion of $H_{int}$ is,
\begin{eqnarray}
	H_{int}=\sum_{k=0}^{\infty} \sum_{m_k=-k}^k \int & & d\vec{r'}
	\frac{1}{2k+1} \delta\rho(\vec{r'})
	\frac{r_<^k}{r_>^{k+1}} \\ \nonumber & & \times 
	Y_{k,m_k}(\theta,\phi)
	Y^{\ast}_{k,m_k}(\theta',\phi').
\end{eqnarray}
where $Y_{k,m_k}$ are the spherical harmonics, $r_< \equiv min(r,r')$ and $r_> \equiv max(r,r')$.
Since the excited electron is practically exterior to
the ion, $r_>$ can be identified with its coordinate and
$r_<$ with the ion electrons. 
The interaction Hamiltonian is hence a sum of terms acting separately on the excited electron and on the ion,
\begin{eqnarray} \label{eq:simplified multipole}
	H_{int} \simeq \sum_{k=0}^{\infty} \sum_{m_k=-k}^k & & \frac{1}{2k+1}
	\frac{1}{r^{k+1}} Y_{k,m_k}(\theta,\phi) 
	\\ \nonumber & & \times \int d\vec{r'}r'^k 
	\delta\rho(\vec{r'}) Y^{\ast}_{k,m_k}(\theta',\phi').
\end{eqnarray}

Eq. (\ref{eq:simplified multipole}) is essentially an expansion 
in the ratio between the characteristic radius of $\delta\rho(\vec{r})$ ($\sim
a_0$) and the 
average radius of the excited electron ($\sim n^2 a_0$). This ratio is small due to the large spatial 
extent of the
excited electron, which follows directly from the softness of the confining
atomic potential at large distances. Roughly, we expect the $k^{th}$ 
term in the sum to
scale like $ \sim 1/n^{2(k+1)}$. Thus, high order terms in the multipole expansion diminish quickly. 

Calculation of matrix elements of the simplified form of $H_{int}$
(\ref{eq:simplified multipole}) involves separate calculations of matrix
elements for
the excited electron, $ \langle n_i,l_i,m_i |\frac{1}{r^{k+1}} Y_{k,m_k}|n_f,l_f,m_f \rangle $, and for the ion, $ \int d\vec{r'}r'^k Y^{\ast}_{k,m_k}(\theta',\phi') \langle \alpha_i|\delta\rho(\vec{r'}) | \alpha_f \rangle $.
The infinite sums in (\ref{eq:simplified multipole}) practically contain only  few terms, due to angular momenta selection rules.  The sum over $m_k$ reduces to a single term with $m_k=m_f-m_i$. As a result of parity conservation, 
the sum over $k$ has non-vanishing terms for either odd or even values of $k$
(but not for both). The $k=0$ term (monopole) vanishes since it is already included in $H_0$.
Due to angular momenta addition rules for the excited electron, $k$ is upper bounded by $l_i +l_f$. The terms in the expansion 
decrease with $k$, so the dipole term is dominant (unless it vanishes). 

Next, we examine the extent to which the interaction couples
degenerate eigenstates of $H_0$.
The odd terms in the multipole expansion, and in particular the dominant 
dipole term 
($k=1$), do not couple degenerate states. This
follows from the observation that terms with odd $k$ contribute only when
the parity of $|\alpha_i \rangle$ and $|\alpha_f \rangle$ are different, and ionic states of opposite parity are non-degenerate (apart from rare cases of accidental degeneracy). 
$H_{int}$ may have non-vanishing matrix elements within a degenerate
sub-space originating from the even $k$ terms in the multipole expansion,
prominently the quadrupole term ($k=2$).
In these cases, one should use degenerate perturbation theory. However, since the quadrupole matrix elements are negligible compared with the dipole contribution, we focus on the latter using non-degenerate perturbation
theory.
 
Consider the matrix elements of the excited electron. In cases where the final state can be described as an ion and a hydrogenic 
electron, just as we assume for
the initial state, the matrix element is calculated by substituting  the appropriate 
hydrogenic wavefunctions and performing the integrals. This is done in appendix
\ref{sec-hydrogen integrals}.
As expected, the matrix elements decrease with increasing k, roughly as $ \sim n^{-2(k+1)}$. Moreover, the matrix elements decrease rapidly as the difference between $n_i$ and $n_f$ increases, because the radial integral is strongly suppressed by the phase difference between the initial and final hydrogenic wavefunctions.

When the final state, $|n_f,l_f,m_f \rangle $, is not a hydrogen-like state, the
calculation of the matrix element depends on the specific atom and particularly on the 
 potential at $r \sim a_0$. However, for such final states, the matrix
elements are negligible because of the small spatial overlap between the confined 
final wavefunction and the relatively extended initial wavefunction\cite{footnote5}. 

To examine the dominant dipole term ($k=1$) we exploit the commutation relation, $[h_e,Y_{1,m}]$. 
On one hand,
\begin{eqnarray}
 	\lefteqn{\langle n_i,l_i,m_i |[h_e,Y_{1,m}]|n_f,l_f,m_f \rangle} & & 
	\\ \nonumber & = & \left( \epsilon_i-\epsilon_f \right)
 	\langle n_i,l_i,m_i | Y_{1,m}|n_f,l_f,m_f \rangle,
\end{eqnarray}
while on the other hand,	
\begin{eqnarray} 
	\lefteqn{ \langle n_i,l_i,m_i |[h_e,Y_{1,m}]|n_f,l_f,m_f \rangle} &  &
	\\ \nonumber &  = & \frac{\hbar^2}{2m}
	\langle n_i,l_i,m_i |[\frac{\vec{l}^2}{r^2},Y_{1,m}]|n_f,l_f,m_f \rangle
	\\ \nonumber  & = & e^2 a_0\frac{l_i(l_i+1)-l_f(l_f+1)}{2}
	\\ \nonumber & & \times
	\langle n_i,l_i,m_i |\frac{Y_{1,m}}{r^2}|n_f,l_f,m_f \rangle \nonumber,
\end{eqnarray}
leading to,
\begin{eqnarray} 
	\lefteqn{\langle n_i,l_i,m_i |\frac{Y_{1,m}}{r^2}|n_f,l_f,m_f \rangle} &
	 \\
	& =\frac{2}{e^2 a_0}
	\frac{\epsilon_i-\epsilon_f}
	{l_i(l_i+1)-l_f(l_f+1)}
	\langle n_i,l_i,m_i |Y_{1,m}|n_f,l_f,m_f \rangle \nonumber.
\end{eqnarray}
Thus, the dipole term vanishes unless $l_f=l_i\pm 1$.
For hydrogen-like wavefunctions we find,
\begin{eqnarray} \label{eq:hydrogen me}
	\lefteqn{\langle n_i,l_i,m_i |\frac{Y_{1,m}}{r^2}|
	n_f,l_i \pm 1,m-m_i \rangle} & & \\
	& \simeq  & \frac{1}{a_0^2} \frac{1}
	{n_i^3 l_i}
	\langle l_i,m_i |Y_{1,m}|l_i\pm1,m-m_i \rangle
	\\ \nonumber & & \times
	\int dr \: u_{n_i,l_i}(r) u_{n_f,l_i\pm1}(r) \nonumber.
\end{eqnarray}
The overlap integral and hence the matrix element is strongly suppressed 
when $n_f$ differs from $n_i$.

An exact calculation of the ionic part of the interaction matrix elements is notably
more difficult, since it depends on the ionic many-body
wavefunctions. Due to parity conservation, the dipole term in 
(\ref{eq:simplified multipole}) has only non-diagonal 
matrix elements, 
\begin{eqnarray} 
	\lefteqn{\int d\vec{r} \: r Y^{\ast}_{1,m}(\theta,\phi)
	\langle \alpha_i|\delta\rho(\vec{r}) | \alpha_f \rangle}&  \\
	& =\langle \alpha_i|\int d\vec{r} \: r Y^{\ast}_{1,m}(\theta,\phi)
	\rho(\vec{r}) | \alpha_f \rangle=\langle \alpha_i|d^1_m |
	\alpha_f \rangle.\nonumber
\end{eqnarray}
where $\{d^1_m\}_{m=\pm1,0}$ is the tensorial representation of the dipole
operator, $\vec{d}=\int d\vec{r} \: \vec{r} \rho(\vec{r})$. From the
Wigner-Eckart theorem we learn that $\vec{d}$ connects only states with equal
$S$ and $M_S$,  $\Delta L=\pm1,0$, $\Delta M_L=\pm1,0$ and opposite parity.

We now estimate the dipole matrix elements $| \langle \alpha_i|\vec{d}| \alpha_f \rangle|^2 $ for final ionic states that appear in the golden rule sum which is restricted by energy conservation. The energy transferred by 
the excited electron ( $< E_i$) is too small for exciting core electrons. Therefore, the  final ionic states that should be summed over involve  valence electrons excitations only.

The dipole operator can be expressed in terms of single particle creation and annihilation operators,
\begin{equation}
	\vec{d}=\sum_{i, j} \vec{d}_{i j} c^{\dagger}_i c_j,
\end{equation}
where $\vec{d}_{i j}$ are the matrix elements of the single particle dipole operator. This sum can be divided as follows,
\begin{equation}
	\vec{d}=\sum_{i,j \in  open \; shells} \vec{d}_{i j} c^{\dagger}_i c_j
	+ \sum_{i\; or  \;j \in core}  \vec{d}_{i j} c^{\dagger}_i c_j .
	\label{eq:ddipole}
\end{equation}
We define a projected dipole operator, $ \displaystyle \vec{d}_{proj}= \sum_{i,j \in open \; shells} \vec{d}_{i j} c^{\dagger}_i c_j$.
For final ionic states with no core excitations, $|\alpha_f \rangle $, 
\begin{equation}
	\langle \alpha_f| \vec{d}|\alpha_i\rangle = 
\langle \alpha_f| \vec{d}_{proj}|
	\alpha_i\rangle,
\end{equation}
and, 
\begin{equation}
	\sum_f | \langle \alpha_f| \vec{d}_{proj}| \alpha_i \rangle |^2 = 
	\langle \alpha_i | \vec{d}_{proj}^2 | \alpha_i \rangle.
\end{equation}
The last equation can be used to estimate the average squared dipole matrix element appearing in the golden rule,
\begin{equation}
	\overline{ | \langle \alpha_f |\vec{d}|\alpha_i \rangle |^2} \sim
	\frac{\langle \alpha_i | \vec{d}_{proj}^2 | \alpha_i \rangle}{N_{tot}},
\end{equation}
where $N_{tot}$, as defined earlier, is the number of final ionic states coupled by the interaction to the initial state.

Similarly, we define a projected density operator, $ \displaystyle \rho_{proj}=\sum_{i,j \in open \; shells} c^{\dagger}_i c_j $. With this definition,
\begin{eqnarray}
	 \vec{d}_{proj}^2 & = & \int d\vec{r} \: d\vec{r'} \: \vec{r} 
	\cdot \vec{r'}
	\rho_{proj}(\vec{r}) \rho_{proj}(\vec{r'}) \\ \nonumber
	&=&  \int d\vec{r} \: \vec{r}^2 \rho_{proj}(\vec{r}) + \int 
	d\vec{r} \: d\vec{r'} \: \vec{r} \cdot \vec{r'} g(\vec{r},\vec{r'}),
\end{eqnarray}
where we have introduced the pair correlation operator $g(\vec{r},\vec{r'}) 
\equiv \rho_{proj}(\vec{r}) \rho_{proj}(\vec{r'})-\delta (\vec{r}-\vec{r'})
\rho_{proj}(\vec{r}) $. 

Thus,
\begin{eqnarray}
	\langle \alpha_i | \vec{d}_{proj}^2 | \alpha_i \rangle & = &
	\int d\vec{r} \: \vec{r}^2 \langle \alpha_i |\rho_{proj}(\vec{r})
	| \alpha_i \rangle \\ \nonumber & &  + \int
        d\vec{r} \: d\vec{r'} \: \vec{r} \cdot \vec{r'} \langle \alpha_i |
	g(\vec{r},\vec{r'}) |\alpha_i \rangle.
\label{eq:d2a}
\end{eqnarray}
The first integral in (\ref{eq:d2a}) is due to auto-correlation while the second integral reflects correlations between electrons. 
For a system of many electrons interacting repulsively, one expects the 
ground state to be a Fermi-liquid-like state, with a short range pair correlation function $g({\vec r},{\vec r}')$ (on the order of the average distance between electrons). Furthermore, since  repulsive interaction tends to keep the valence electrons away from one another, $g({\vec r},{\vec r}')<0$ for small $|{\vec r}-{\vec r}'|$. Consequently, 
\begin{equation} 
         \langle \alpha_i | \vec{d}_{proj}^2 | \alpha_i \rangle 
	\stackrel{\sim}{<}
        \langle \alpha_i | \int d\vec{r} \: \vec{r}^2 \rho_{proj}(\vec{r})
        |\alpha_i \rangle .
\end{equation}                    

Alternatively, $\langle \alpha_i| \vec{d}_{proj}^2 |\alpha_i \rangle $ can be estimated using the chaotic nature of the many-body eigenstates. $|\alpha_i\rangle$ can be decomposed into a sum of single Slater determinants, $|\alpha_i\rangle= \sum_m \Delta_
m |\delta_m\rangle $, where $ \Delta_m \equiv \langle \delta_m | \alpha_i \rangle$. Following Flambaum\cite{Flambaum} we assume that $|\alpha_i \rangle$ are chaotic superpositions, so the coefficients $\Delta_m $ are random. The number of principle components in the decomposition is denoted by ${\cal N}$, so that $|\Delta_m| \sim \frac{1}{\sqrt{{\cal N}}} $.
Thus,
\begin{eqnarray} \label{eq:chaotic}
	\langle \alpha_i | \vec{d}_{proj}^2 | \alpha_i \rangle &=& 
	\sum_{m,n} \Delta_m^* \Delta_n \langle \delta_m | \vec{d}_{proj}^2 |
	 \delta_n \rangle \\ \nonumber
	&=& \sum_m |\Delta_m|^2 \langle \delta_m | \vec{d}_{proj}^2 |
         \delta_m \rangle 
	\\ \nonumber & & + \sum_{m \neq n} 
	 \Delta_m^* \Delta_n \langle \delta_m | \vec{d}_{proj}^2 |
         \delta_n \rangle.
\end{eqnarray}
It is easily shown that the matrix elements of $ \vec{d}_{proj}^2 $ in  a basis of single Slater determinants are all $O(N_{eff})$, where $N_{eff}$ is the number of valence electrons. The first sum in (\ref{eq:chaotic}) is $O(N_{eff})$, since it has ${\cal N}$ non-negative terms of 
order $\frac{1}{{\cal N}} O(N_{eff}) $. The second sum is of the same order, since it is a random sum of ${\cal N}^2 $ terms of order $\frac{1}{{\cal N}} O(N_{eff}) $.

Both points of view lead, then, to the estimate,
\begin{equation}  \label{eq:ion me} 
	\overline{ | \langle \alpha_f |\vec{d}|\alpha_i \rangle |^2} \sim
	\frac{N_{eff} R_{val}^2}{N_{tot}},
\end{equation}
where $ R_{val}$ is the average radius of a valence electron. 
The number of valence electrons, as well as their average radius, vary between atoms due to the shell structure.

A very rough estimate of the interaction matrix element, taking into account
only
the dominant dipole contribution (for $|i\rangle$ and $|f\rangle$ satisfying all
the
selection rules described above), is obtained by combining (\ref{eq:hydrogen
me}) and (\ref{eq:ion me}),
\begin{eqnarray}  \label{eq:dipole me}
	|\langle i|H_{int}|f\rangle|^2 & \sim &
	\left(\frac{1}{3}\right)^2 \: \left(\frac{1}{a_0^2 n_i^3 l_i}\right)^2
	\: \frac{e^4 N_{eff} R_{val}^2}{N_{tot}}
	\\ \nonumber & \sim &\frac{1}{N_{tot}}
	\left(\frac{e^2}{a_0} \right)^2
	\frac{N_{eff} R_{val}^2}{10 (n_i^3 l_i)^2 },
\end{eqnarray}
for $n_f$ close to $n_i$. 
The dependence upon the specific atom
under consideration comes from the ionic part of the matrix
element via $N_{tot},N_{eff}$ and $R_{val}$. The
dependence of the matrix elements on the initial  excitation energy
 is mainly due to the excited electron.  Since $N_{tot}$ grows exponentially with $N_{eff}$, the interaction matrix elements are very small compared with the level spacing of the single particle Hamiltonian, $h_e$.

\subsection{{\it e-e} Lifetime}

\subsubsection{Alkali Atom} \label{sec-alkali}

The considerations above show that
in alkali atoms below the ionization threshold
there is no level broadening due to {\it e-e} interaction, and 
each low lying many-body eigenstate can be identified 
with a particular single-particle excitation 
(i.e. there is a large overlap $\stackrel{<}{\sim} 1$
between these two states). 
The number of single particle excitations in this energy range is very large (infinite in principle), because the single particle spectrum becomes hydrogen-like as the ionization threshold is approached. The existence of many single particle excitations 
with no {\it e-e} broadening in alkali atoms follows from the non-uniform single particle density of states, characterized by a gap  at the Fermi energy and becoming dense near the ionization threshold.

\begin{figure}[!htb] 
	\centerline{\psfig{figure=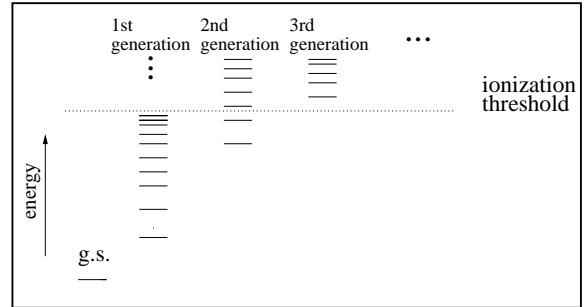,width=3in}}
	\caption{A schematic illustration of the hierarchical tree for an 
	alkali atom.}
	\label{fig-alkali atom tree}
\end{figure}

As mentioned earlier, the problem of {\it e-e} lifetime of single particle excitations
can be mapped onto the problem of localization on a hierarchical tree. Let us consider 
the hierarchal tree of an alkali atom shown schematically in figure
 (\ref{fig-alkali atom tree}). Any $n$-particle $n$-hole excitation with $n>1$
involves excitation of $n-1$ electrons from  closed shells, and therefore a
large excitation energy.  
This is
manifested in the tree by a large gap between the lowest state in the $n^{th}$
generation and the ground state, a gap that grows rapidly with $n$
 (see fig. \ref{fig-alkali atom tree}).
The structure of the hierarchical tree forces low lying single particle 
excitation to be localized in Fock space. Therefore, they are not 
broadened by {\it e-e}
interaction. 
The suppression of {\it e-e} broadening is characteristic of systems with a gap at the Fermi level, and it persists to excitation energies considerably larger than the gap. The atomic system is unique in that the single particle spectrum becomes dense above the energy gap, so the number of single particle excitations in alkali atoms for which {\it e-e} broadening is suppressed is very large.

\subsubsection{The Complex Atom} \label{sec-complex}

We argued earlier that the density of states of ionic excitations in complex atoms is exponentially large. 
The question is whether this density of states is sufficient to
induce broadening of single particle excitations. The general criteria for
applying the golden rule to the calculation of lifetime is that the resulting scattering
rate, $\hbar/ \tau_{ee} \simeq \overline{|\langle H_{int} \rangle |}^2 G_{final}(E)$ (where
$\overline{|\langle H_{int} \rangle |}$ is the average interaction matrix
element and
$G_{final}(E)$ is the relevant density of states) is larger than the level spacing of the relevant
final unperturbed states $1/G_{final}(E)$. This requirement is equivalent to the condition,
\begin{equation} \label{eq:condition gr}
	\frac{1}{G_{final}(E)}< \overline{|\langle H_{int} \rangle |}.
\end{equation} 
$G_{final}(E)$  refers only to final states which are coupled to the initial state by the interaction, and 
generally may be considerably smaller than the full many-body 
density of states.
In our case, $G_{final}(E)$ is smaller
than the full many-body density of states due to the selection rules associated with symmetry (for the dominant dipole term). 
Nevertheless,  $G_{final}(E)$ still grows exponentially with the number of active
electrons $N_{eff}$. 
$N_{tot}$ is the number of accessible final states,
and  the relevant density of states is roughly given by $G_{final}(E) \simeq N_{tot}a_0/e^2$.
We saw in the previous section that 
$\overline{|\langle H_{int} \rangle |} \propto 1/\sqrt{N_{tot}}$. 
Since $1/G_{final}(E) \propto 1/N_{tot}$, eventually for large enough
$N_{tot}$,  (\ref{eq:condition gr}) is satisfied. We conclude that {\it for
large
enough $N_{tot}$, {\it e-e} interaction leads to finite lifetime of single particle
excitations}. 

The lifetime of a single particle 
excitation $|i\rangle$ with energy $E$ is,
\begin{equation}  \label{eq:gr for atoms}
	\frac{1}{\tau_{ee}}=\frac{2 \pi}{\hbar} \sum_{n_f,l_f,m_f}
	g_{ion}(E+\frac{e^2}{2 a_0 n_f^2}) \overline{|\langle
	i| H_{int}|f \rangle|^2}.
\end{equation}
The summation is restricted to final states of the excited electron, as we have averaged the interaction matrix elements over the ionic states and
correspondingly introduced the ionic density of states.
Taking into account only dipole matrix elements (eq. \ref{eq:dipole me}), 
one obtains,
\begin{eqnarray}
	\frac{1}{\tau_{ee}} & \sim & \frac{2 \pi}{\hbar} 
	\left(\frac{N_{tot} a_0}{e^2} \right) \left( \frac{e^4}{N_{tot} a_0^2}
	\frac{N_{eff} R_{val}^2 }{10 (n_i^3 l_i)^2} \right)
	\\ \nonumber & \sim &
	\frac{2 \pi}{\hbar} \frac{e^2}{a_0}
	\frac{N_{eff} R_{val}^2}{10 (n_i^3 l_i)^2}. 
\end{eqnarray}
The single particle excitation spectrum is hydrogen-like with $E_{n_i} \sim -e^2/2 a_0 n_i^2$, so the single particle level spacing is $\propto 1/n_i^3$. 
For large enough $n_i$, the
level spacing of the excited electron becomes smaller than the level spacing of
the ion. For such single particle excitations, no {\it e-e} lifetime
can be defined and (\ref{eq:gr for atoms}) becomes irrelevant.  

The calculated {\it e-e} scattering rate depends on the specific atom through $N_{eff} R_{val}^2$, which depends irregularly on $Z$ due to the shell structure. 
For all atoms ($Z<105$), the number of valence electrons $< 14$, and the resulting {\it e-e} scattering rate is much smaller than the single particle level spacing. 

One may consider atoms with larger $Z$.
According to the TF model (see \ref{sec-sp TF} above), $N_{eff}
\propto Z^{1/3}$ for small $Z$, but in the limit  $Z \rightarrow \infty $ it saturates to a constant. This yields {\it e-e} scattering rates which are independent of $Z$ in the limit $Z \rightarrow \infty $. The TF model, however, ignores the shell structure. 
The shell structure suggests that the largest open sub-shell in an atom may have $Z^{1/3}$ electrons, so the number of valence electrons fluctuates, as a function of $Z$, between $O(1)$ and $O(Z^{1/3})$ electrons. 
The average radius of the valence electrons, $R_{val}$, also depends on $Z$. As the number of valence electrons increases, we expect $R_{val}$ to decrease because the screening of the nucleus is less efficient.
This tends to reduce the fluctuations in $N_{eff} R_{val}^2 $ as a function of $Z$.

Thus, {\it e-e} scattering rates in atoms fluctuate with $Z$. 
A small fraction of atoms have $O(Z^{1/3})$ valence electrons. For such atoms the typical scattering rates increase as a function of $Z$ ($1/\tau_{ee} \propto Z^{\alpha}, \: \alpha<1/3 $), and eventually, for large enough $Z$, become larger than the single particle level spacing.
For most atoms, however, the number of valence electrons is small and the {\it e-e} scattering rates of single particle excitations remain smaller than the single particle level spacing. 

For a specific complex atom, the width of single particle excitations decreases 
as the excitation energy increases, in contrast with the analogous width in quantum dots. 
This decrease is a direct consequence of the soft atomic potential 
at large distances ($r \gg a_0$). A small increase in the excitation energy 
 amounts to a substantial increase of the average radius of the excited electron.
 As a result, the interaction matrix elements and therefore the {\it e-e} broadening 
are significantly reduced.

\section{Conclusions} \label{sec-con}

We estimated the {\it e-e} scattering rate of single particle
excitations in atoms below the ionization threshold. We conclude that
{\it e-e} interaction does not limit the number of discrete single particle
excitations observable below the ionization threshold in naturally occurring atoms, since the  broadening it induces is small
compared with the level spacing of single particle excitations.  In practice, broadening due to electron-photon
interaction limits the number of resolved single particle excitations. 
We find that {\it e-e} scattering rate in
atoms decreases with excitation energy (below the ionization
threshold), contrary to its behavior in Fermi liquids.

The {\it e-e} scattering rates in ballistic (clean) and diffusive (dirty) QDs have
been calculated theoretically\cite{sivan}. These calculations indicate that {\it e-e} interaction severely limits the 
number of
observable levels in QDs. In a ballistic QD containing $N$ electrons, only 
$\sim  \sqrt{N}$ discrete single particle resonances can be resolved. The 
effect of {\it e-e} interaction is even more dramatic in diffusive QDs. The number 
of resolved single particle resonances in a 0-D diffusive QD\cite{footnote7} is approximately equal to the dimensionless conductance 
$g$ which is typically $\stackrel{<}{\sim} 10-20$. In addition, the {\it e-e} scattering rate of single 
particle
excitations in QDs is shown to increase as the excitation energy becomes
larger. This Fermi-liquid like behavior is in contrast with our results for 
atoms.

The profound difference of the {\it e-e} scattering rate  in QDs and in atoms can be attributed to
the different confining potentials. In atoms, the confining potential is very soft
at large distances $\sim 1/r$, while in QDs, the potential has
sharper boundaries $\sim r^2$. The characteristics of the confining potential
determine the spatial extent of the single particle wavefunctions as well as the
single particle spectrum. These properties have substantial influence on  {\it e-e} scattering rates. 

In QDs the confining potential generates  a fairly uniform single particle
spectrum in the vicinity of the Fermi level. The 
single particle spectrum leads to a relevant density of states (in
this case, the 2-electron 1-hole density of states) which grows quadratically with excitation energy. The resulting {\it e-e} scattering
rates increase with excitation energy. The spectrum consists of
a
small number of discrete single particle resonances, beyond which the broadening exceeds the single particle level spacing, and the single 
particle resonances merge and form a continuous spectrum.

In contrast, the Coulomb confining potential in atoms produces a dense single particle excitation spectrum as the ionization threshold is approached. The spatial
extent of an excited electron's wavefunction increases rapidly with excitation
energy, due to the softness of the Coulomb potential at large distances. 
The interaction matrix elements hence decrease with excitation energy leading to reduced {\it e-e} scattering rates.
The number of observed discrete levels is found not to be limited by {\it e-e} interaction.

\acknowledgements 
This research was supported by a grant from the US--Israel Binational
Science Foundation, by the Israeli Academy of Sciences and by the German-Israeli foundation DIP. A.S. is supported by the V. Ehrlich career development
chair. 

\appendix
\section{}  \label{sec-hydrogen integrals}

The matrix element of $Y_{k,m_k}(\theta,\phi)/r^{k+1}$ between two hydrogenic
wavefunctions $|n,l,m \rangle$ and $|n',l',m'\rangle$ encountered in section
(\ref{sec-me interaction}) is 
calculated in the following way. The radial and angular integrations are separated,
\begin{eqnarray} 
	\lefteqn{\langle n',l',m'|  Y_{k,m_k}(\theta,\phi)\frac{1}{r^{k+1}}
	|n,l,m \rangle } \\ \nonumber
	& &= \int_0^{\infty} dr \frac{1}{r^{k+1}} u_{n',l'}(r) 
	u_{n,l}(r) 
	\\ & & \times \int d\omega Y_{l',m'}^{\star}(\theta,\phi)
	Y_{k,m_k}(\theta,\phi) Y_{l,m}(\theta,\phi).\nonumber 
\end{eqnarray}
The angular integral is expressed in terms of Wigner's 3-j symbol,
\begin{eqnarray} \label{eq:integral 3Ylm}
	\lefteqn{\int d\omega Y_{l',m'}^{\star}(\theta,\phi)
	Y_{k,m_k}(\theta,\phi) Y_{l,m}(\theta,\phi)} & &  \\ \nonumber
	& = & \sqrt{\frac{(2l'+1)(2k+1)(2l+1)}{4 \pi}} 
	\\ \nonumber & & \times
	\left( \begin{array}{ccc}
		l' & k & l \\
		0 & 0 & 0
		\end{array} \right)  
	\left( \begin{array}{ccc}
		l' & k & l \\
		-m' & m_k & m
		\end{array} \right)  .
\end{eqnarray}
The radial integral gives,
\begin{eqnarray} 
	\lefteqn{\int_0^{\infty} dr \frac{1}{r^{k+1}} u_{n',l'}(r)
	u_{n,l}(r)} 
	\\ \nonumber
	& = & \int_0^{\infty} dr \frac{1}{r^{k+1}} 
	\sqrt{\frac{(n-l-1)!}{a_0 n^2(n+l)!}} e^{-r/na_0} \left(\frac{2r}{n a_0}
	\right)^{l+1} 
	\\ \nonumber & & \times
	L_{n-l-1}^{2l+1}(2 r/na_0) 
	\sqrt{\frac{(n'-l'-1)!}{a_0 n'^2(n'+l')!}} e^{-r/n'a_0}
	\\ \nonumber & & \times  \left(\frac{2r}
	{n' a_0}\right)^{l'+1}L_{n'-l'-1}^{2l'+1}(2 r/n'a_0) 
	\\ \nonumber & = &
	\frac{2^{l+l'+2}}{n^{l+2}n'^{l'+2} a_0^{k+1}} \sqrt{\frac{(n-l-1)!}
	{(n+l)!} \frac{(n'-l'-1)!}{(n'+l')!}}
	\\ \nonumber & & \times
	\sum_{i=0}^{n-l-1}\sum_{j=0}^{n'-l'-1} (-1)^{i+j}  \left( \stackrel{
	\textstyle n+l}{n-l-1-i} \right)\nonumber 
	\\ \nonumber & & \times 
	\left( \stackrel
	{\textstyle n'+l'}{n'-l'-1-j}\right) 
	\frac{1}{i!} \left( \frac{2}{n}
	\right)^i \frac{1}{j!} \left( \frac{2}{n}\right)^j
	\\ \nonumber & & \times
	\frac{(l+l'-L+1+i+j)!}{(\frac{1}{n}+\frac{1}{n'})^{l+l'-L+2+i+j}}.
\end{eqnarray}


\begin{thebibliography}{99}

\bibitem{footnote-1} The subject of the present paper was posed by the late Arkadi Aronov in the form of the following question ``What is the difference between an Uranium atom and a quantum dot?''

\bibitem{sivan} U.Sivan, F.P.Milliken, K.Milkove, S.Rishton, Y.Lee, J.M.Hong,
V.Boegli, D.Kern and M. DeFranza,  Europhys. Lett. {\bf 25}, 605 (1994),
U.Sivan, Y.Imry and A.G.Aronov, Europhys. Lett.
{\bf 28}, 115 (1994).

\bibitem{footnote0} We note the analogy between our calculation and  the calculation of the
lifetime of an atomic level due to electron-photon interaction in a box of
volume $V$. There, the interaction matrix elements are $\propto 1/\sqrt{V}$, 
while the
density of states is $\propto V$. The
volume in this calculation plays a role similar to $N_{tot}$ in our 
calculation -  for large enough volume, the golden rule 
is valid and the resulting lifetime is independent of $V$.

\bibitem{AAGK} B.L.Altshuler, Y.Gefen, A.Kamenev and L.S.Levitov, Phys. Rev. Lett. {\bf 78}, 2803 (1997).

\bibitem{footnote1} Excitations above the ionization threshold are coupled 
by the free electromagnetic field to a continuous spectrum of an ion and an
unbound electron, leading to auto-ionization. Auto-ionization processes dominate
the {\it e-e} lifetime above the threshold.

\bibitem{Flambaum} V.V.Flambaum, A.A.Gribakina, G.F.Gribakin, M.G.Kozlov, Phys.
Rev. A {\bf 50(1)}, 267 (1994).

\bibitem{LL} L.D.Landau and E.M.Lifshitz, {\em Quantum Mechanics} 
(Pergamon, New York, 1975).

\bibitem{Latter} R.Latter, Phys. Rev. {\bf 99(2)}, 510 (1955).

\bibitem{March} N.H.March, J. Chem. Phys. {\bf 76(3)}, 1430 (1982).

\bibitem{footnote2} This
continuous density of states would 
be the result of averaging $g(\epsilon)=\sum_i
\delta(\epsilon-\epsilon_i)$ over energy scales larger than the single particle
level spacing.

\bibitem{footnote3} The exact value of the bound
is meaningless since it depends on the definition of the relevant energy range.
The important point is the existence of an upper bound independent of $Z$.

\bibitem{atomic calc} F.Herman and S.Skillman, {\em 
Atomic Structure Calculations} (Prentice-Hall, Inc., New Jersey, 1963).

\bibitem{footnote4} We do not specify explicitly the spin quantum number since 
it is unaffected by the interaction.
 
\bibitem{footnote5} The 
Pauli exclusion principle limits the excited electron
to single particle states unoccupied by the ion electrons. The matrix elements related to transitions from the initial highly excited 
state to an occupied single particle states are negligible, and the
neglect of the Pauli principle should not significantly affect our result for  {\it e-e} scattering rates.

\bibitem{footnote7} By 0-D 
we mean $L<\sqrt{\frac{\hbar D}{\epsilon}}$, where $L$ is the system's linear 
dimension, $D$ the diffusion coefficient and $\epsilon$ the single particle 
excitation energy.


\end{thebibliography}
\end{document}